\documentclass{article}
\usepackage[T1]{fontenc}
\usepackage[utf8]{inputenc}
\usepackage[]{ismir} 
\usepackage{amsmath,cite,url}
\usepackage{graphicx}
\usepackage{color}

\usepackage[dvipsnames]{xcolor}

\usepackage{enumitem}

\title{High-Resolution Sustain Pedal Depth Estimation from Piano Audio Across Room Acoustics}

\multauthor
  {Kun Fang$^{1, 2,*}$ \hspace{0.8cm} Hanwen Zhang$^{1, 2,*}$ \hspace{0.8cm} Ziyu Wang$^{3, 4}$ \hspace{0.8cm} Ichiro Fujinaga$^{1,2}$}
  {
   $^{1}$ McGill University\quad 
   $^{3}$ New York University\quad
   $^{4}$ Music X Lab, MBZUAI\\
   $^{2}$ The Centre for Interdisciplinary Research in Music Media and Technology (CIRMMT)
  }

\def\authorname{K. Fang, H. Zhang, Z. Wang and I. Fujinaga}

\begin{document}

\maketitle
\makeatletter
\renewcommand\thefootnote{\fnsymbol{footnote}} 
\footnotetext[1]{Authors with equal contribution}
\makeatother
\renewcommand\thefootnote{\arabic{footnote}}
\setcounter{footnote}{0} 

\begin{abstract}
Piano sustain pedal detection has previously been approached as a binary on/off classification task, limiting its application in real-world piano performance scenarios where pedal depth significantly influences musical expression. This paper presents a novel approach for high-resolution estimation that predicts continuous pedal depth values. We introduce a Transformer-based architecture that not only matches state-of-the-art performance on the traditional binary classification task but also achieves high accuracy in continuous pedal depth estimation. Furthermore, by estimating continuous values, our model provides musically meaningful predictions for sustain pedal usage, whereas baseline models struggle to capture such nuanced expressions with their binary detection approach. Additionally, this paper investigates the influence of room acoustics on sustain pedal estimation using a synthetic dataset that includes varied acoustic conditions. We train our model with different combinations of room settings and test it in an unseen new environment using a ``leave-one-out'' approach. Our findings show that the two baseline models and ours are not robust to unseen room conditions. Statistical analysis further confirms that reverberation influences model predictions and introduces an overestimation bias.
\end{abstract}

\section{Introduction}\label{sec:introduction}


The use of the sustain pedal is a highly personal and indispensable aspect of piano performance. In treatises \cite{gieseking2013piano,neuhaus2008art}, legendary piano teachers Karl Leimer (1858–1944) and Heinrich Neuhaus (1888–1964) both placed great emphasis on recognizing the sound shaped by the sustain pedal, where small variations in timing and depth can create dramatic effects in tones. The sound effect created by the sustain pedal is also highly sensitive to environmental acoustics. Professional pianists adjust pedaling meticulously in realtime to achieve their desired tonal quality, and it is the interplay between pedaling and external acoustic factors that shapes the listening experience. In fact, in earlier days, it was a suggested practice to simulate sustain pedaling effects using artificial reverberation algorithms as noted in \cite{lehtonen_analysis_2007}. 

However, in music information retrieval (MIR), many studies have adopted the term ``sustain pedal detection'' and approached this problem as a binary (on/off) classification \cite{liang_transfer_2019, liang_piano_2019, kong_high-resolution_2021, yan_skipping_2021, yan_scoring_2024}, overlooking the subtleties of intermediate states.
Recent research \cite{kong_high-resolution_2021, yan_skipping_2021, yan_scoring_2024} also primarily uses the MAESTRO dataset \cite{hawthorne_enabling_2019}, which is currently one of the largest available collections of paired audio and MIDI files recorded on Yamaha Disklaviers (acoustic grand pianos with high-precision MIDI capture and playback systems). Unfortunately, the room conditions of MAESTRO's audio recordings are entirely unknown and also potentially unpredictable. This raises further questions about the generalizability of these models across different acoustic environments when trained solely on MAESTRO, in addition to the oversimplification inherent in binary classification.

To this end, we redefine the task as continuous sustain pedal depth estimation, framing it as a regression problem. We propose a Transformer-based model with a conventional structure that includes a convolutional layer for feature extraction, followed by an n-layer Transformer Encoder. Additionally, we incorporate a mixed evaluation strategy, combining F1 of classification across various bin thresholds with continuous mean absolute error (MAE). Our results demonstrate that our continuous approach outperforms binary models both quantitatively and qualitatively when treating baseline binary predictions as continuous values while maintaining comparable performance in strict binary classification scenarios.


Moreover, we use MAESTRO’s aligned MIDI performances and synthesize a dataset with diverse acoustic environments to evaluate the robustness of pedal detection algorithms. 
In our experiments, we observe that both our model and the baseline models experience a performance drop when trained on real recordings and tested on synthetic audio rendered in different room acoustics. To further investigate this issue, we conduct a leave-one-out experiment using synthetic data from multiple acoustic environments, which reveals that the model struggles to generalize to unseen acoustics. In addition, statistical analysis shows that models trained solely on real recordings tend to produce higher pedal predictions as reverberation increases. This bias is reduced when the model is trained on a more diverse dataset in terms of room acoustics.
These findings demonstrate the influence of room acoustics on model predictions of the sustain pedal depth values.
Our contributions are summarized as follows:
\begin{enumerate}[noitemsep]
    \item \textbf{Continuous high-resolution pedal depth estimation.}  
    We reformulate sustain pedal detection as a continuous-valued regression task and propose a Transformer-based model. Our method matches state-of-the-art performance in binary classification setting and outperforms baselines in multi-class and regression metrics.

    \item \textbf{Robustness analysis under varied room acoustics.}  
    We conduct controlled leave-one-out experiments showing that unseen room conditions degrade model performance. Statistical analysis reveals a consistent overestimation bias in reverberant settings, which is reduced through acoustic diversity in training.

    \item \textbf{A synthetic dataset for controlled acoustic generalization research.}  
    We introduce a new dataset rendered from MAESTRO's MIDI files with multiple room configurations, supporting systematic studies of model robustness and the effects of room acoustics on pedal detection.
\end{enumerate}


\section{Related Work}

Two important early studies from an acoustic perspective examined how sustain pedal affects sound. The first one \cite{lehtonen_analysis_2007} found that pressing the pedal increases decay times in mid-range frequencies and alters the tone through sympathetic resonance. The further study \cite{lehtonen_analysis_2009} investigated part-pedaling and identified three phases: initial free vibration, damper-string interaction, and final free vibration. This provides scientific evidence that varying the pedal depth allows for gradual and continuous control over the sound rather than a simple on/off function.

In MIR-related field, Liang et al. implemented an optical sensor \cite{liang_piano_2017} and explored the relationship between pedaling techniques and physical gestures. Their subsequent work includes classifying isolated notes with different sustain pedal timings \cite{liang_detection_2017}, classifying short excerpts from a few recordings with four levels of pedal depth \cite{liang_measurement_2018}, and detecting legato-pedal onsets by measuring sympathetic resonance \cite{liang_piano_2018}. These studies have shown potential but have not yet fully solved the pedal depth estimation problem in real-world scenarios.
However, while moving towards deep-learning approach, Liang et al. shifted their focus to binary pedal detection in \cite{liang_transfer_2019, liang_piano_2019} with a large collection of synthetic data.
Kong et al. applied the best CNN in \cite{liang_piano_2019} to real recordings from the MAESTRO dataset, achieving a 0.791 F1 score \cite{kong_high-resolution_2021}.

Some recent comprehensive piano transcription systems have also integrated sustain pedal detection and achieved state-of-the-art binary pedal detection accuracy\cite{kong_high-resolution_2021,yan_skipping_2021,yan_scoring_2024}. These approaches employed the same binary labeling of sustain pedals following \cite{liang_transfer_2019,liang_piano_2019} (split pedal on/off at 63/64 MIDI CC64 values) during training. The current state-of-the-art binary pedal detection has an activation-level F1 score of 0.954\cite{yan_scoring_2024} on MAESTRO.

\section{Methodology}


This section introduces the data representation and model architecture for frame-wise continuous pedal depth estimation. We first describe how input features and training targets are derived from audio and aligned MIDI data. Then, we present the Transformer-based model, including a convolutional block and multi-task prediction output layers. We detail the multi-objective loss function for learning both fine-grained pedal depth and discrete change events. Finally, we describe a synthetic dataset rendered under multiple acoustic environments to enable systematic evaluation of model generalization and robustness.


\begin{figure*}[htbp]
\centering
\includegraphics[width=1.0\linewidth]{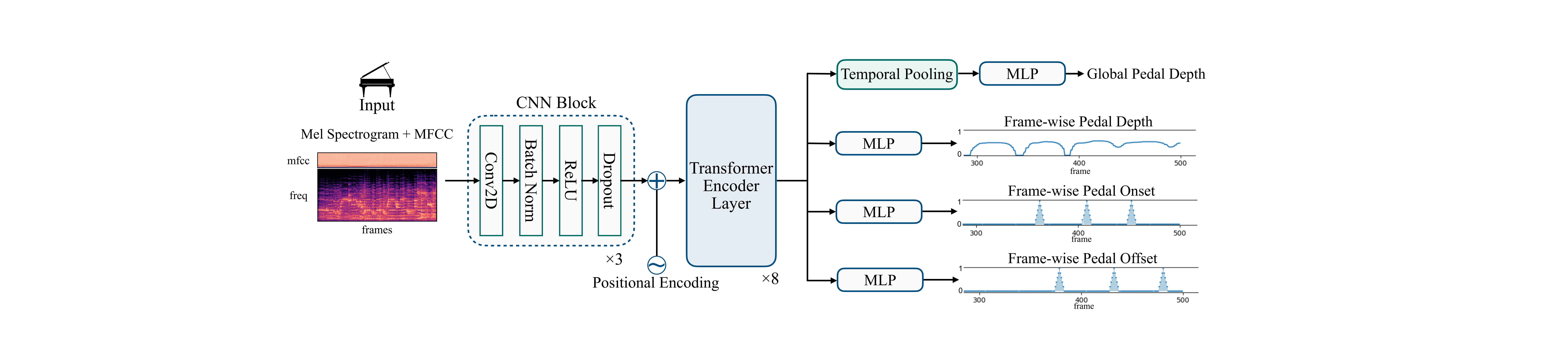}
\caption{Overview of the model architecture. Input features are first processed by a convolutional block, followed by a Transformer Encoder that captures temporal dependencies. Four prediction heads (MLPs) output frame-wise pedal depth, onset, offset, and a global pedal depth value.}
\label{fig:model_architecture}
\end{figure*}

\subsection{Data Representation}

Each music piece is segmented into 500-frame clips using a sliding window, with each segment corresponding to approximately 5 seconds of audio. The input audio is resampled to 16 kHz, which is sufficient to capture the highest piano note (C8 at 4186 Hz), as noted by Kong et al.~\cite{kong_high-resolution_2021}. A log-mel spectrogram with 229 frequency bands is computed using a short-time Fourier transform with a Hann window length of 2048 samples and a hop length of 160 samples, resulting in a frame rate of 100 frames per second. In addition, we extract 20 MFCC coefficients. 
All feature extraction is performed using librosa\footnote{\textit{librosa} 0.10.1, https://doi.org/10.5281/zenodo.8252662.}.

Each segment passed to the model has shape $[T, F]$ where $T=500$ is the number of frames and $F=249$ is the feature dimension. As shown in Figure~\ref{fig:model_architecture}, the model predicts and learns from four types of targets derived from aligned MIDI data:
\begin{itemize}
    \item \textbf{Frame-Wise Pedal Depth}: A continuous sequence of length $T$, with each value normalized to the range $[0, 1]$ by dividing the MIDI CC64 values (range 0--127) by 127.

    \item \textbf{Frame-Wise Pedal Onset}: A binary sequence of length $T$, where value 1 indicates a pedal activation event (pedal pressed down) in the current frame.

    \item \textbf{Frame-Wise Pedal Offset}: A binary sequence of length $T$, where a value of 1 indicates a pedal release event (pedal lifted) in the current frame.
    
    \item \textbf{Global Pedal Depth}: A single scalar representing the average pedal level within the segment. It summarizes the overall pedaling behavior and provides a global supervision target. 
\end{itemize}

We follow the event-based definition of pedal onsets and offsets in~\cite{kong_high-resolution_2021}, where onsets are triggered by a rising edge in the CC64 curve and offsets by a falling edge. To improve learning and reduce sensitivity to label noise, we adopt the same strategy as~\cite{kong_high-resolution_2021} by generating \textit{soft labels} around the onset and offset events, shown in Figure~\ref{fig:model_architecture}. This approach improves convergence during training and enables the model to better localize pedal change boundaries. This representation design allows the model to capture both the continuous control of the sustain pedal and the discrete pedal changes. 

\subsection{Model Architecture}

The overall model architecture is illustrated in Figure~\ref{fig:model_architecture}. We follow a common design used in temporal prediction tasks such as beat tracking~\cite{Zhao2022,Hung2022,Foscarin2024}, consisting of a convolutional block, a Transformer encoder, and multiple prediction layers. The convolutional block includes three 2D convolutional layers with batch normalization, ReLU activation, and max pooling along the frequency axis, which compress the spectral information and extract high-level representations from the log-mel input. The resulting sequence is then fed into an 8-layer Transformer encoder with a hidden size of 256, 8 attention heads, and a feed-forward dimension of 1024. The Transformer models temporal dependencies across frames. On top of the encoder, we apply four separate MLPs to predict frame-wise pedal depth, pedal onset, pedal offset, and a global pedal depth value. The frame-wise outputs have the same length as the input sequence (500 frames), while the global prediction is computed by applying mean pooling over the encoded sequence. A dropout rate of 0.15 is used during training.

\subsection{Loss Functions}

Our model is trained using a multi-task objective that combines continuous regression and binary classification losses. This design reflects the structure of the prediction targets, which include both continuous-valued pedal depth curve and discrete pedal change events. The total loss function is defined as:
\begin{equation}\label{loss}
\mathcal{L}_{\text{total}} = \lambda_1 \mathcal{L}_{\text{pedal}} + \lambda_2 \mathcal{L}_{\text{global}} + \lambda_3 \mathcal{L}_{\text{onset}} + \lambda_4 \mathcal{L}_{\text{offset}}
\end{equation}
\noindent where $\mathcal{L}_{\text{pedal}}$ is the frame-wise mean squared error (MSE) between the predicted and ground truth continuous pedal depth, $\mathcal{L}_{\text{global}}$ is the MSE loss for predicting the global average pedal depth across the segment, and $\mathcal{L}_{\text{onset}}$ and $\mathcal{L}_{\text{offset}}$ are binary cross-entropy (BCE) losses applied to the predicted onset and offset event sequences, respectively.
MSE is chosen for $\mathcal{L}_{\text{pedal}}$ and $\mathcal{L}_{\text{global}}$ because the prediction targets are continuous values, and minimizing squared error encourages the model to match the fine-grained temporal structure of pedaling gestures. In contrast, BCE is used for $\mathcal{L}_{\text{onset}}$ and $\mathcal{L}_{\text{offset}}$ because they represent binary event classification tasks at the frame level. This multi-objective setup allows the model to simultaneously learn both low-level control signals (pedal depth curves) and high-level timing events (onset and offset boundaries). The scalar weights $\lambda_1$ through $\lambda_4$ are manually tuned to balance the contribution of each objective.

\subsection{Dataset Synthesis Across Environments}
To explore how the same pedaling (and playing) might sound given different acoustic conditions, we used MAESTRO v3\cite{hawthorne_enabling_2019} and synthesized their recorded MIDI performances using Pianoteq 8 Stage.\footnote{Pianoteq is a complete physical modeling software developed at Institute of Mathematics of Toulouse, France. https://www.modartt.com/} We selected four combinations of varying reverb parameters to simulate distinct acoustic environments. The specific configurations are summarized in Table~\ref{tab:room_piano_settings}. These parameters are mostly default room presets in Pianoteq. We output audio with the standard stereo microphone setup at 44.1 kHz, 16-bit. 





\begin{table*}[ht]
\small
\centering
\begin{tabular}{c p{1.8cm} p{0.8cm} p{0.7cm} p{0.7cm} p{0.8cm} p{0.8cm} p{0.8cm} p{0.8cm} p{4.0cm}}
\hline
\textbf{Room} & \textbf{Name} & \textbf{Mix} & \textbf{Dur} & \textbf{Size} & \textbf{PreD} & \textbf{D.Mix} & \textbf{D.Amt} & \textbf{D.FB} & \textbf{Piano Model} \\
\hline
1 & Dry Room      & -      & -    & -    & -     & -    & -    & -    & NY Steinway D Classical \\

2 & Clean Studio  & +10dB  & 0.4s & 12m  & 0s    & 6\%  & 60ms & 0\%  & NY Steinway D Classical \\

3 & Concert Hall  & +50dB  & 4s   & 50m  & 0.01s & 15\% & 60ms & 5\%  & NY Steinway D Classical \\

4 & Church        & +10dB  & 2.5s & 18m  & 0s    & 25\% & 30ms & 0\%  & Bösendorfer 280VC Classical \\

\hline
\end{tabular}
\caption{
Summary of room acoustic settings and piano models used for synthesized audio. Each room configuration specifies the reverb mix level (Mix), reverb duration (Dur), room size (Size), pre-delay (PreD), delay mix (D.Mix), delay amount (D.Amt), delay feedback (D.FB), and the piano model used for synthesis. No reverberation is applied in the dry room. In the rest of this paper, we refer to these rooms using their index (e.g., Room~1 for Dry Room.).
}
\label{tab:room_piano_settings}
\end{table*}
\section{Experiments}

\begin{table*}[ht]
\small
\centering
\small
\begin{tabular}{l|ccc|ccc|cc}
\hline
\textbf{Model} & \multicolumn{3}{c|}{\textbf{Binary}} & \multicolumn{3}{c|}{\textbf{4-Class}} & \textbf{MSE} $\downarrow$ & \textbf{MAE} $\downarrow$\\
\cline{2-7}
               & P $\uparrow$ & R $\uparrow$ & F1 $\uparrow$
& P $\uparrow$ & R $\uparrow$ & F1 $\uparrow$ &  &  \\
\hline
Kong et al. \cite{kong_high-resolution_2021} & 0.9374 & 0.9374 & 0.9373 & 0.4752  & 0.6786 & 0.5588 & 0.0762 & 0.1579 \\
Yan and Duan \cite{yan_scoring_2024} & \textbf{0.9446} & \textbf{0.9445} & \textbf{0.9443} & 0.4790  & 0.6836  & 0.5631  & 0.0710  & 0.1528 \\
Ours (BCE) & 0.8907 & 0.8911 & 0.8906 & 0.6039 & 0.6507 & 0.6180 & 0.0584 & 0.1537 \\
Ours & 0.8975 & 0.8971 & 0.8973 & \textbf{0.6849}  & \textbf{0.6971}  & \textbf{0.6863}  & \textbf{0.0425} & \textbf{0.1339}\\

\hline
\end{tabular}
\caption{
Comparison of model performance on binary and 4-class classification, along with regression metrics. Baselines support only binary classification, but we computed their 4-class scores for comparison. Our model demonstrates lower MSE and MAE, along with significantly higher 4-class scores. Precision, recall, and F1 scores remain consistent across the 4-class results. This is achieved by predicting continuous pedal depth values beyond binary classification.}
\label{tab:quantitative_results}
\end{table*}


This section evaluates our model's performance in pedal depth estimation and compares it with baseline models. We show that our model has advantages in both quantitative and qualitative analysis.


\subsection{Training}

Our model contains approximately 7.2 million parameters. Training is conducted on a single Tesla V100-SXM2-32GB GPU with a batch size of 16. We use the AdamW optimizer \cite{adam} with a learning rate of $5 \times 10^{-4}$. Each training epoch takes around 5 hours and 45 minutes to complete. We trained the model for a total of 48 hours, observing the best checkpoint near 192,000 steps in epoch 9, around which the metrics appeared to indicate convergence. The total loss in Equation~\ref{loss} is optimized with the following weights: $\lambda_1 = 0.6$ for frame-wise pedal depth, $\lambda_2 = 0.2$ for global pedal depth, $\lambda_3 = 0.1$ for pedal onset, and $\lambda_4 = 0.1$ for pedal offset.

\subsection{Baselines}

We use the models from \cite{kong_high-resolution_2021,yan_scoring_2024} as our baselines, as they are, to the best of our knowledge, the state-of-the-art piano transcription models that include sustain pedal detection. Both models take log-mel spectrograms as input. 
Kong et al. \cite{kong_high-resolution_2021} trained a separate CRNN model for sustain pedal detection using the weighted sum of binary cross-entropy (BCE) losses for sustain pedal onset, offset and value, 
while Yan and Duan \cite{yan_scoring_2024} put sustain pedal events along with note events and performed transcription as a whole using a Transformer Encoder and semi-CRF. 

We also include an additional version of our model, denoted as ``Ours (BCE)'' in Table~\ref{tab:quantitative_results} and \ref{tab:robustness_baseline_results}, which is designed for comparison and ablation purposes. It uses the same architecture as our main model but adopts BCE loss for all four objectives.


\subsection{Quantitative Evaluation}
Comparing our model with baseline methods poses a challenge because the baselines treat sustain pedal prediction as a binary classification task and are trained using binary cross-entropy loss. In contrast, our model is explicitly designed to estimate a continuous pedal depth curve by regressing values within the range $[0, 1]$. We acknowledge that this fundamental difference in approach limits direct comparability between methods.

To enable fair comparison, we first discretize our continuous predictions to match the frame-wise binary labels used by the baselines. We also compute frame-wise mean squared error (MSE) and mean absolute error (MAE) for the baselines, treating their binary outputs as approximations of continuous values, though we recognize this may not reflect their intended use case. Finally, to assess the resolution of continuous predictions, we introduce an accuracy-with-tolerance metric that evaluates prediction correctness under varying error thresholds. The evaluation metrics are summarized as follows:
\begin{itemize}
    \item \textbf{Frame-Wise F1 Score}: Evaluates classification performance by discretizing predictions into binary and multi-class labels. While baseline models are trained only for binary output, we compute multi-class results for all models.
    
    \item \textbf{MSE and MAE}: Measure the accuracy of predicted vs. ground truth continuous pedal values.
    \item \textbf{Accuracy with Tolerance}: Evaluates the model’s precision and resolution along the continuous pedal depth axis. A prediction is considered correct if its absolute error falls within a specified threshold.
\end{itemize}

The quantitative results in Table~\ref{tab:quantitative_results} show that our model performs comparably to the baselines in binary classification, while achieving better results in a continuous pedal prediction scenario with lower MSE and MAE. 
Furthermore, in piano pedagogy and performance studies, the basic discrete categorization of sustain pedal depth usually involves quarter, half, three-quarters, and full pedal as described in \cite{schnabel1954modern} and \cite{banowetz_pianists_1985}. When evaluated on this musically meaningful 4-class classification task, our model achieves a 0.6863 F1 score while both baselines score around 0.56, suggesting potential of our continuous approach.

\begin{figure}[ht]
    \centering
    \includegraphics[width=0.8\linewidth]{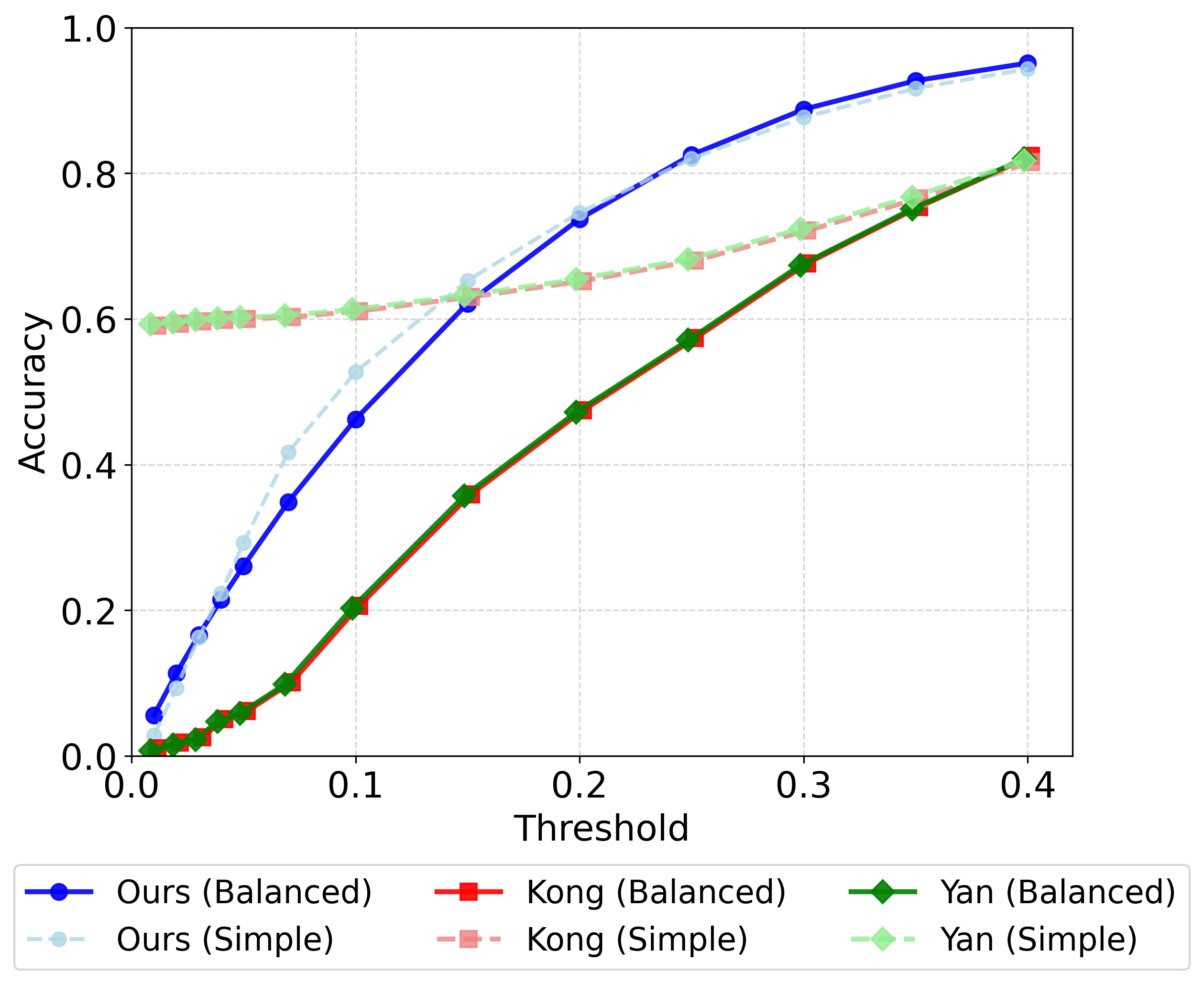}  
     \caption{Balanced accuracy and simple average accuracy of our model compared to baselines (Kong et al.\cite{kong_high-resolution_2021}; Yan and Duan\cite{yan_scoring_2024}) across thresholds ranging from 0.01 to 0.4.}
    \label{fig:tolerance_accuracy}
\end{figure}

\begin{figure*}[!ht]
    \centering
    \includegraphics[width=1.01\linewidth]{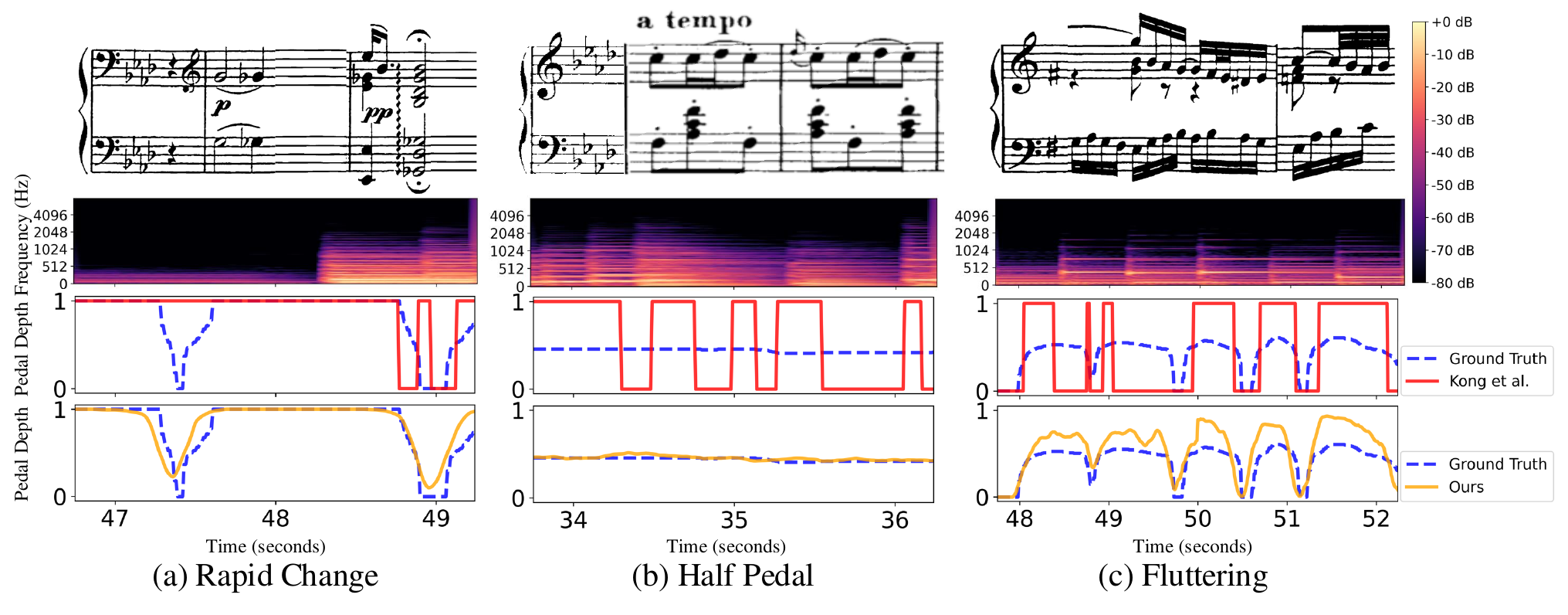}
    \caption{Comparison of our predicted pedal curves and the baseline binary model \cite{kong_high-resolution_2021} against ground truth pedal depth values, plotted alongside aligned mel-spectrograms and their respective scores: (a) F. Chopin (1810–1849) \textit{Polonaise-Fantaisie} Op. 61, mm. 6–7; (b) F. Schubert (1797–1828) \textit{Impromptu in F Minor}, Op. 142 No. 4, mm. 45–46; and (c) J. S. Bach (1685–1750) \textit{Prelude and Fugue in E Minor}, BWV 855, mm. 11–12. }
    \label{fig:qualitative_plot}
\end{figure*}

We introduce \textit{balanced accuracy with tolerance} to evaluate model performance at different resolution levels while accounting for the heavy data imbalance toward extreme pedal values, common in piano playing. This metric is particularly useful in real-world scenarios when small prediction errors are acceptable. We calculate the \textit{balanced accuracy with tolerance} as
$A_{\text{weighted}} = \left(\sum_{i=0}^{N-1} w_i \cdot a_i\right)/\left(\sum_{i=0}^{N-1} w_i\right)$ where weight $w_i$ is determined by the inverse square root of the frequency of each bin, normalized by the total sum of inverse square root frequency of all bins. We set the number of frequency bins as $N=128$, matching the discrete 0-127 ground truth CC64 values. As in {Figure~\ref{fig:tolerance_accuracy}, our model achieves higher balanced accuracy than the baselines, starting from tight thresholds below 0.05 and improving further with looser tolerances. Simple accuracy is more sensitive to imbalance, but our model surpasses the baselines from threshold 0.15 onward, indicating that most prediction errors are relatively small.



\subsection{Qualitative Analysis}
In this section, we present example segments from the test set featuring various expressive sustain pedal gestures, shown in Figure~\ref{fig:qualitative_plot}, including gradual half-pedaling, flutter-pedaling, and rapid pedal changes within very short durations. These examples are common and representative in the classical piano repertoire. 

Example (a) shows legato pedaling with rapid full pedal changes, where the baseline either misses changes or predicts excessive oscillations. Example (b) is a half-pedaling instance where a binary classification algorithm fails to capture the intermediate pedal depth. In the flutter-pedaling region in example (c), our model identifies repeated partial presses and releases, whereas the baseline reduces these expressive variations to abrupt binary changes. 
These examples indicate that our method can model real pedaling behavior more effectively and generate musically meaningful pedal curves, providing a more faithful representation of pianists’ precise control over the pedal depth compared to previous binary prediction methods.

\begin{figure*}[!ht]
    \centering
    \includegraphics[width=0.9\textwidth]{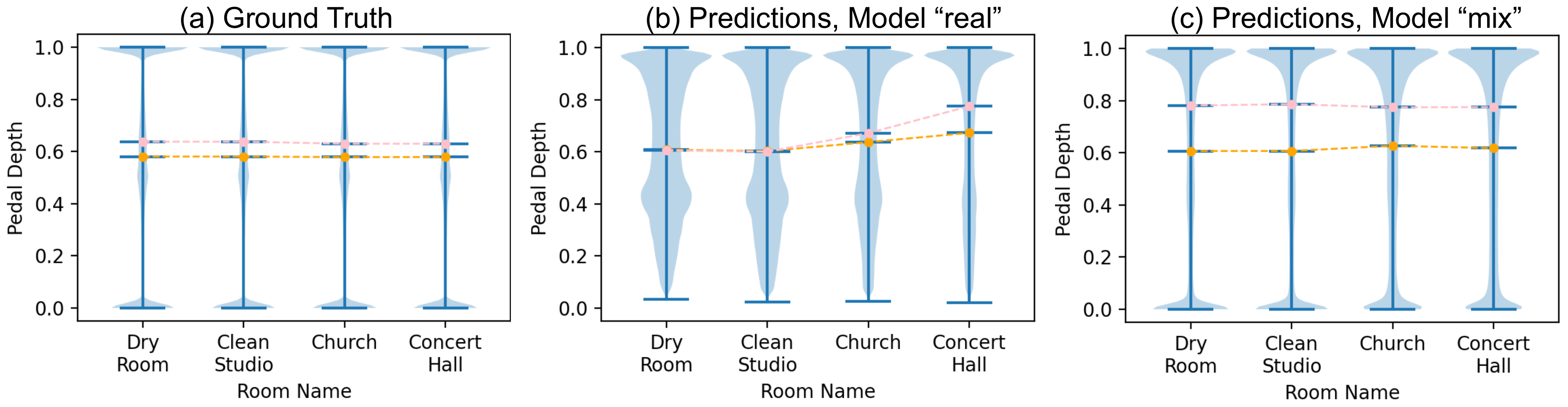}
    \caption{
    Violin plots showing the distribution of predicted pedal depth values (b and c) and ground truth (a) across four room settings: Dry Room (R1), Clean Studio (R2), Church (R4), and Concert Hall (R3). (b) is our model trained solely on real data, while (c) learns from real data + training data from R1, R2, and R3. Orange dashed lines indicate the mean pedal depth across rooms, and pink dashed lines mark the median.
    }
    \label{fig:violinplot_comparison}
\end{figure*}

\section{Pedal Prediction Robustness}

We observe that both baseline models, as well as our own, struggle to generalize from real recordings to synthetic data. To address this, we conduct a leave-one-out experiment showing that training on a more diverse set of room conditions improves our model’s ability to generalize to unseen acoustic environments. Further analysis supports our claim that \textit{current sustain pedal detection models are not robust to changes in room acoustics; moreover, their predictions are significantly influenced by reverberation, revealing a strong sensitivity to acoustic environments.}


\subsection{Robustness Analysis}
To evaluate the model’s robustness to acoustic variations, we follow the original MAESTRO dataset split for our synthetic data. If a piece belongs to the training set, all its synthetic versions are also in training. This ensures that the model never encounters the musical content of test data during training regardless of room conditions.
We first evaluate two baselines and our model on synthetic test sets and report mean absolute error (MAE) in Table~\ref{tab:robustness_baseline_results}. Although all models perform similarly on in-domain data, the baselines degrade notably on out-of-domain (OOD) sets, indicating limited generalization. Our model shows the same trend. To investigate further, we conduct leave-one-room-out experiments by training on various combinations of synthetic data. As shown in Table~\ref{tab:robustness_results}, the model performs worst on any room it was not trained on, confirming the importance of acoustic diversity for model generalization.

\begin{table}[ht]
\small
\centering
\setlength{\tabcolsep}{4pt} 
\begin{tabular}{@{}cccccc@{}}
\hline
& \multicolumn{5}{c}{Test Data} \\
\cline{2-6}
Model & O & R1 & R2 & R3 & R4 \\
\hline
Kong et al. \cite{kong_high-resolution_2021} 
& 0.1579 & 0.1773 & 0.1780 & 0.2458 & 0.2247 \\
Yan and Duan\cite{yan_scoring_2024} 
& 0.1528 & 0.2217 & 0.2195 & 0.2449 & 0.2365 \\
Ours (BCE) & 0.1537 & \textbf{0.1654} & \textbf{0.1681} & 0.2091 & \textbf{0.1894} \\
Ours 
& \textbf{0.1339} & 0.1889 & 0.1841 & \textbf{0.2039} & 0.1905 \\
\hline
\end{tabular}
\caption{Mean absolute error (MAE)$\downarrow$ of baseline models tested on the original MAESTRO recordings (O) and synthetic audio with settings R1, R2, R3, and R4, respectively.}
\label{tab:robustness_baseline_results}
\end{table}
\begin{table}[ht]
\small
\centering
\setlength{\tabcolsep}{4pt} 
\begin{tabular}{ccccc}
\hline
& \multicolumn{4}{c}{Test Data} \\
\cline{2-5}
Training Data & R1 & R2 & R3 & R4 \\
\hline
R2+R3 & {\setlength{\fboxsep}{1pt}\fbox{0.1125}} & 0.1119 & 0.1453 & {\setlength{\fboxsep}{1pt}\fbox{0.1654}} \\
R1+R3 & 0.1107 & {\setlength{\fboxsep}{1pt}\fbox{0.1145}} & 0.1433 & {\setlength{\fboxsep}{1pt}\fbox{0.1713}} \\
R1+R2 & 0.0979 & 0.0981 & {\setlength{\fboxsep}{1pt}\fbox{0.1863}} & {\setlength{\fboxsep}{1pt}\fbox{0.1779}} \\
R1+R2+R3 & 0.1062 & 0.1065 & 0.1439 & {\setlength{\fboxsep}{1pt}\fbox{0.1619}} \\
\hline
\end{tabular}
\caption{MAE$\downarrow$ of our model trained with original MAESTRO audio and synthetic audio with settings (rooms) R2+R3, R1+R3, R1+R2, and R1+R2+R3, respectively, tested on rooms R1, R2, R3, and R4. R4 is never included in any training configuration and serves as a held-out condition for evaluating generalization to unseen acoustic environments. Out-of-domain (OOD) results are boxed.}
\label{tab:robustness_results}
\end{table}
\subsection{Effect of Room Acoustics on Pedal Prediction}


To further understand how room acoustics influence frame-level pedal prediction, 
we analyze the distribution of predicted pedal depth values across four synthetic room conditions ordered by non-linear progression of reverb levels.


Figure~\ref{fig:violinplot_comparison} presents violin plots comparing ground truth pedal values, predictions from our model trained only on real recordings, and predictions from our model trained on both real and synthetic audio rendered under R1, R2, and R3. The ground truth distributions remain stable across rooms, with consistent ranges, medians, and symmetric shapes, which is unsurprising since all samples are synthesized from the same MIDI CC64 values. In contrast, the real-data-only model shows a noticeable upward shift in both median and distribution shape as reverberation increases, suggesting that the model interprets increasing reverberant energy as more use of the sustain pedal. The mixed-data model (right plot) produces distributions that more closely match the ground truth, with mean values and overall shapes remaining consistent across rooms and no clear upward trend as reverb increases.

These findings reveal that generalization is a key challenge in sustain pedal detection. Models trained on a single acoustic setting tend to produce biased predictions when tested in new environments. Our analysis confirms that room acoustics can significantly affect model behavior. However, we also show that incorporating training data from multiple acoustic settings helps mitigate this issue, suggesting that acoustic diversity is a promising direction for improving model robustness.

\section{Conclusion}

This paper presents a high-resolution sustain pedal estimation model capable of predicting continuous pedal depth values beyond binary on/off states.\footnote{Available at https://github.com/kunfang98927/PedalDetection.} Through quantitative and qualitative evaluations, we show that our model enables more detailed analysis of expressive piano performance by capturing finer-grained continuous pedaling depth values, while maintaining performance comparable to SOTA baselines on previous binary detection tasks. We acknowledge the inherent limitations when comparing continuous and binary classification approaches. We further evaluate our model’s robustness across varying acoustic conditions using ``leave-one-out'' test with different acoustic settings. Additionally, the robustness analysis reveals that reverberation systematically increases prediction error and introduces an overestimation bias in pedal depth value predictions. These findings highlight the importance of modeling continuous pedal depth values and accounting for acoustic variability in sustain pedal detection tasks.

\bibliography{ISMIR2025_template}

\begin{thebibliography}{10}
\providecommand{\url}[1]{#1}
\csname url@samestyle\endcsname
\providecommand{\newblock}{\relax}
\providecommand{\bibinfo}[2]{#2}
\providecommand{\BIBentrySTDinterwordspacing}{\spaceskip=0pt\relax}
\providecommand{\BIBentryALTinterwordstretchfactor}{4}
\providecommand{\BIBentryALTinterwordspacing}{\spaceskip=\fontdimen2\font plus
\BIBentryALTinterwordstretchfactor\fontdimen3\font minus \fontdimen4\font\relax}
\providecommand{\BIBforeignlanguage}[2]{{%
\expandafter\ifx\csname l@#1\endcsname\relax
\typeout{** WARNING: IEEEtran.bst: No hyphenation pattern has been}%
\typeout{** loaded for the language `#1'. Using the pattern for}%
\typeout{** the default language instead.}%
\else
\language=\csname l@#1\endcsname
\fi
#2}}
\providecommand{\BIBdecl}{\relax}
\BIBdecl

\bibitem{gieseking2013piano}
W.~Gieseking and K.~Leimer, \emph{{Piano Technique}}.\hskip 1em plus 0.5em minus 0.4em\relax Courier Corporation, 2013.

\bibitem{neuhaus2008art}
H.~Neuhaus, \emph{{The Art of Piano Playing}}.\hskip 1em plus 0.5em minus 0.4em\relax Kahn and Averill, 2008.

\bibitem{lehtonen_analysis_2007}
H.-M. Lehtonen, H.~Penttinen, J.~Rauhala, and V.~Välimäki, ``Analysis and modeling of piano sustain-pedal effects,'' \emph{The Journal of the Acoustical Society of America}, vol. 122, no.~3, pp. 1787--1797, September 2007.

\bibitem{liang_transfer_2019}
B.~Liang, G.~Fazekas, and M.~B. Sandler, ``Transfer learning for piano sustain-pedal detection,'' in \emph{2019 International Joint Conference on Neural Networks ({IJCNN})}.\hskip 1em plus 0.5em minus 0.4em\relax Budapest, Hungary: {IEEE}, 2019, pp. 1--6.

\bibitem{liang_piano_2019}
------, ``Piano sustain-pedal detection using convolutional neural networks,'' in \emph{Proceedings of {IEEE} International Conference on Acoustics, Speech and Signal Processing ({ICASSP})}, Brighton, United Kingdom, 2019, pp. 241--245.

\bibitem{kong_high-resolution_2021}
Q.~Kong, B.~Li, X.~Song, Y.~Wan, and Y.~Wang, ``High-resolution piano transcription with pedals by regressing onset and offset times,'' \emph{{IEEE}/{ACM} Transactions on Audio, Speech and Language Processing}, vol.~29, pp. 3707--3717, October 2021.

\bibitem{yan_skipping_2021}
Y.~Yan, F.~Cwitkowitz, and Z.~Duan, ``Skipping the frame-level: Event-based piano transcription with neural semi-{CRFs},'' in \emph{Advances in Neural Information Processing Systems}, vol.~34, Virtual, 2021, pp. 20\,583--20\,595.

\bibitem{yan_scoring_2024}
Y.~Yan and Z.~Duan, ``Scoring time intervals using non-hierarchical transformer for automatic piano transcription,'' in \emph{Proceedings of the International Society for Music Information Retrieval Conference ({ISMIR})}, San Francisco, United States, 2024.

\bibitem{hawthorne_enabling_2019}
C.~Hawthorne, A.~Stasyuk, A.~Roberts, I.~Simon, C.-Z.~A. Huang, S.~Dieleman, E.~Elsen, J.~Engel, and D.~Eck, ``Enabling factorized piano music modeling and generation with the {MAESTRO} dataset,'' in \emph{Proceedings of the 7th International Conference on Learning Representations}, New Orleans, Louisiana, United States, 2019.

\bibitem{lehtonen_analysis_2009}
H.-M. Lehtonen, A.~Askenfelt, and V.~Välimäki, ``Analysis of the part-pedaling effect in the piano,'' \emph{The Journal of the Acoustical Society of America}, vol. 126, no.~2, pp. EL49--EL54, July 2009.

\bibitem{liang_piano_2017}
B.~Liang, G.~Fazekas, M.~B. Sandler, and A.~{McPherson}, ``Piano pedaller: A measurement system for classiﬁcation and visualisation of piano pedalling techniques,'' in \emph{Proceedings of The International Conference on New Interfaces for Musical Expression ({NIME})}, Copenhagen, Denmark, 2017, pp. 325--329.

\bibitem{liang_detection_2017}
B.~Liang, G.~Fazekas, and M.~B. Sandler, ``Detection of piano pedaling techniques on the sustain pedal,'' in \emph{Proceedings of Audio Engineering Society Convention 143}, New York, United States, 2017.

\bibitem{liang_measurement_2018}
------, ``Measurement, recognition, and visualization of piano pedaling gestures and techniques,'' \emph{Journal of the Audio Engineering Society}, vol.~66, no.~6, pp. 448--456, June 2018.

\bibitem{liang_piano_2018}
------, ``Piano legato-pedal onset detection based on a sympathetic resonance measure,'' in \emph{Proceedings of the 26th European Signal Processing Conference ({EUSIPCO})}, Rome, Italy, 2018.

\bibitem{Zhao2022}
J.~Zhao, G.~Xia, and Y.~Wang, ``Beat transformer: Demixed beat and downbeat tracking with dilated self-attention,'' in \emph{Proceedings of the International Society for Music Information Retrieval Conference (ISMIR)}, Bengaluru, India, 2022.

\bibitem{Hung2022}
Y.~Hung, J.~Wang, X.~Song, W.~T. Lu, and M.~Won, ``Modeling beats and downbeats with a time-frequency transformer,'' in \emph{Proceedings of the IEEE International Conference on Acoustics, Speech and Signal Processing (ICASSP)}, Singapore, 2022, pp. 401--405.

\bibitem{Foscarin2024}
F.~Foscarin, J.~Schl{"u}ter, and G.~Widmer, ``Beat this! accurate beat tracking without dbn postprocessing,'' in \emph{Proceedings of the International Society for Music Information Retrieval Conference (ISMIR)}, San Francisco, United States, 2024.

\bibitem{adam}
\BIBentryALTinterwordspacing
I.~Loshchilov and F.~Hutter, ``Decoupled weight decay regularization,'' in \emph{Proceedings of the International Conference on Learning Representations (ICLR)}, New Orleans, Louisiana, United States, 2019. [Online]. Available: \url{https://openreview.net/forum?id=BkgQ1j09K7}
\BIBentrySTDinterwordspacing

\bibitem{schnabel1954modern}
K.~U. Schnabel, \emph{Modern Technique of the Pedal: A Piano Pedal Study}.\hskip 1em plus 0.5em minus 0.4em\relax New York, United States: Mills Music, 1954.

\bibitem{banowetz_pianists_1985}
J.~Banowetz, \emph{The Pianist’s Guide to Pedaling}.\hskip 1em plus 0.5em minus 0.4em\relax Bloomington, {IN}: Indiana University Press, 1985.

\end{thebibliography}

%
%
%
%

\end{document}